\documentclass[amsfonts,amssymb,amsmath,byrevtex,twoside]{revtex4}
\newtheorem{theorem}{Theorem}

\begin{document}

\title{FLUCTUATIONS, CORRELATIONS AND  FINITE VOLUME EFFECTS IN HEAVY ION COLLISION}

\author{Ludwik Turko}
\email{turko@ift.uni.wroc.pl}
\affiliation{Institute of Theoretical Physics, University of Wroc{\l}aw,\\
Pl. Maksa Borna 9, 50-204  Wroc{\l}aw, Poland}

\date{February 23, 2007}

\begin{abstract}
Finite volume corrections to higher moments are important observable quantities.
They make possible to differentiate between different statistical ensembles even in
the thermodynamic limit. It is shown that this property is a universal one. The
classical grand canonical distribution is compared to the canonical distribution in
the rigorous procedure of approaching the thermodynamic limit.
\end{abstract}
\maketitle
\section{Introduction}

Fluctuations and correlations measured in heavy ion collision processes give better
insight into dynamical and kinematical properties of the dense hadronic medium
created in ultrarelativistic heavy ion collisions. Particle production yields are
astonishingly well reproduced by thermal models, based on the assumption of
noninteracting gas of hadronic resonances \cite{brs}. Systems under considerations
are in fact so close to the thermodynamic limit that final volume effects can be
neglected --- at least when productions yields are considered.

The aim of the paper is to show that finite volume effects become more and more
important when higher moments, \emph{e.g.} correlations and fluctuations are
considered. The basic physical characterization of the system described by means of
the thermal model are underlying probability densities that given physical
observables of the system have specified values. The only way to reproduce those
probability distribution is by means of higher and higher probability moments. Those
moments are in fact the only quantities which are phenomenologically available and
can be used for the verification of theoretical predictions. Finite volume effects
are also important for the lattice QCD calculations.

Particle yields in heavy ion collision are the first moments, so
they lead to rather crude comparisons with the model. Fluctuations
and correlations are second moments so they allow for the better
understanding of physical processes in the thermal equilibrium.

A preliminary analysis of the increasing volume effects was given in
\cite{crt1,crt2}. It has been rigorously shown an influence of $\mathcal{O}(1/V)$
terms for a new class physical observables --- semi-intensive quantities
\cite{crt2}. Those results completely explained also ambiguities noted in
\cite{begun}, related to ''spurious non-equivalence'' of different statistical
ensembles used in the description of heavy ion collision processes.

This paper is devoted to  a further analysis of $\mathcal{O}(1/V)$
terms. It is shown that those terms are not specific for systems
with subsidiary internal symmetries but appear also in the simplest
''classical`` problems of statistical physics.

\section{Choice of variables}

In the thermodynamical limit the relevant probabilities
distributions are those related to densities. These distributions
are expressed by moments calculated for densities --- not for
particles. In the practice, however, we measure particles --- not
densities as we do not know related volumes. Fortunately, volumes
can be omitted by taking corresponding ratios.

Let us consider \emph{e.g.} the density variance $\Delta n^2$. This can be written
as
\[\Delta n^2=\langle n^2\rangle - \langle n\rangle^2 =
\frac{\langle N^2\rangle - \langle N\rangle^2}{V^2}\,.\]%
By taking the relative variance
 \[\frac{\Delta n^2}{\langle n\rangle^2}=\frac{\langle N^2\rangle - \langle N\rangle^2}{\langle N\rangle^2}\,,\]
 volume-dependence vanishes.

\subsection{Semi-intensive variables}

A special care should be taken for calculations of ratios of
particles moments. Although moments are extensive variables their
ratios can be finite in the thermodynamic limit. These ratios are
examples of semi-intensive variables. They are finite in the
thermodynamic limit but those limits depend on volume terms in
density probability distributions. One can say that semi-intensive
variables ''keep memory'' where the thermodynamic limit is realized
from.

Let consider as an example the scaled  particle variance
\[\frac{\langle N^2\rangle - \langle N\rangle^2}{\langle N\rangle}=
V\frac{\langle n^2\rangle - \langle n\rangle^2}{\langle n\rangle}\,.\]%

The term
\[\frac{\langle n^2\rangle - \langle n\rangle^2}{\langle n\rangle}\,.\]
tends to zero in the thermodynamic limit as $\mathcal{O}(V^{-1})$. So a behavior of
the scaled particle variance depends on the $\mathcal{O}(V^{-1})$ term in the scaled
density variance. A more detailed analysis of semi-intensive variables is given in
\cite{crt2}.

To clarify this approach let us consider a well known classical problem of Poisson
distribution but taken in the thermodynamic limit.

\section{Grand canonical and canonical ensembles}
\subsection{Poisson distribution in the thermodynamic limit}

Let us consider the grand canonical ensemble of noninteracting gas.
A corresponding statistical operator is
  \begin{equation}\label{stat operator GC-P}
    \hat{D}=\frac{\,e^{-\beta\hat H+\gamma\hat N}}{\,\text{Tr}{\,e^{-\beta\hat H+\gamma\hat N}}}
\end{equation}

This leads to the partition function
\begin{equation}\label{GC-P part fn}
    \mathcal{Z}(V,T,\gamma)=\,e^{z\,e^\gamma}\,.
\end{equation}

where $z$ is  one-particle partition function
\begin{equation}
z(T,V)=\frac{V}{(2\pi)^3}\int d^3p\,\,e^{-\beta E(p)} \equiv V z_0(T)\,,
\end{equation}

A $\gamma$ parameter ($=\beta\mu$) is such to provide the given
value of the average particle number $\langle N\rangle=V\langle
n\rangle$. This means that
\begin{equation}\label{particle factor}
    \,e^{\gamma}= \frac{\langle n\rangle}{z_0}\,.
\end{equation}

Particle moments can be written as
\begin{equation}\label{particle moments}
    \langle N^k\rangle= \frac{1}{\mathcal{Z}}\frac{\partial^k\mathcal{Z}}
    {\partial\gamma^k}\,.
\end{equation}

The parameter $\gamma$ is  taken in final formulae as a function
$\gamma(\langle n\rangle,z_0)$ from Eq \eqref{particle factor}.

The resulting probability distribution to obtain $N$ particles under
condition that the average number of particles is $\langle N\rangle$
is equal to Poisson distribution

\[P_{\langle N\rangle}(N)=\frac{{\langle N\rangle}^N}{N!}\,e^{-\langle N\rangle}\,.\]

We introduce corresponding probability distribution $\mathcal{P}$ for the particle
number density $n=N/V$

\begin{equation}\label{probab dens}
\mathcal{P}_{\langle n\rangle}(n;V)= V P_{V\langle n\rangle}(V n)=V\frac{(V\langle
n\rangle)^{V n}}{\Gamma(V n+1)} \,e^{-V\langle n\rangle}\,.
\end{equation}

 For large $V n$  we are using an asymptotic form of Gamma function

 \[\Gamma(V n+1)\sim\sqrt{2\pi}(V n)^{V n-1/2}\,e^{-V n}\left\{1+\frac{1}{12 V n}+
 \mathcal{O}(V^{-2})\right\}\,.\]

This gives

\begin{equation}\label{prob dens as1}
    \mathcal{P}_{\langle n\rangle}( n;V)\sim V^{1/2}\frac{1}{\sqrt{2\pi n}}
\left(\frac{\langle n\rangle}{ n}\right)^{V n} \,e^{V( n-\langle
n\rangle)}\left\{1-\frac{1}{12 V n}+\mathcal{O}(V^{-2})\right\}
\end{equation}

This expression in singular in the $V\to\infty$ limit. To estimate a large volume
behavior of the probability distribution \eqref{probab dens} one should take into
account a generalized function limit. So we are going to calculate an expression

 \[\langle G\rangle_V=\int dn\, G( n)\mathcal{P}_{\langle n\rangle}(n;V)\,,\]

where $\mathcal{P}_{\langle n\rangle}(n;V)$ is replaced by the asymptotic form from
Eq \eqref{prob dens as1}. In the next to leading order in $1/V$ one should calculate

\begin{equation}\label{Poiss t lim}
    V^{1/2}\frac{1}{\sqrt{2\pi}}\int d n\frac{G( n)}{ n^{1/2}}\,e^{V S( n)} -
V^{-1/2}\frac{1}{12\sqrt{2\pi}}\int d n\frac{G( n)}{ n^{3/2}}\,e^{V S( n)}\,.
\end{equation}

where
\[S( n)= n\ln\langle n\rangle -  n\ln n +  n - \langle n\rangle\,.\]

An asymptotic expansion of the function $\langle G\rangle_V$ is given by the
classical Watson-Laplace theorem

\begin{theorem}
  Let $I=[a,b]$ be the finite interval such that
  \begin{enumerate}
    \item $\max\limits_{x\in I} S(x)$ is reached in the single point $x=x_0$, \mbox{$a<x_0<b$}.
    \item $f(x),S(x)\in C(I)$.
    \item $f(x), S(x)\in C^\infty$ in the vicinity of $x_0$, and $S^{''}(x_0)\neq 0$.
  \end{enumerate}
  Then, for $\lambda\to\infty,\ \lambda\in S_\epsilon$, there is an asymptotic expansion
\begin{subequations}\label{laplace}
  \begin{eqnarray}
    F[\lambda]&\thicksim &\,e^{\lambda S(x_0)}\sum\limits_{k=0}^\infty c_k\lambda^{-k-1/2}\,,
    \label{laplace main}\\
    c_k &=&\frac{\Gamma(k+1/2)}{(2k)!}\left(\frac{d}{dx}\right)^{2k}
\left.\left[f(x)\left(\frac{S(x_0)-S(x)}{(x-x_0)^2}\right)^{-k-1/2}\right]\right\vert_{x=x_0}\,.
\label{laplace coeff}
\end{eqnarray}
\end{subequations}
  $S_\epsilon$~is here a segment $|\arg z|\leqslant\frac{\pi}{2}-\epsilon<\frac{\pi}{2}$ in
the complex $z$-plane.
\end{theorem}

To obtain $\mathcal{O}(1/V)$ formula the first term in \eqref{Poiss
t lim} should be calculated till the next to leading order term in
$1/V$. For the second term it is enough to perform calculations in
the leading order only.

The first term gives the contribution
\begin{subequations}
  \begin{equation}\label{first}
  V^{1/2}\frac{1}{\sqrt{2\pi}}\int d n\frac{G( n)}{ n^{1/2}}\,e^{V S( n)}
   = G(\langle n\rangle)+\frac{1}{12\langle n\rangle V} G(\langle n\rangle)+\frac{\langle n\rangle}{2
    V}G^{''}(\langle n\rangle)\,,
\end{equation}
and the second term gives
  \begin{equation}\label{second}
   V^{-1/2}\frac{1}{12\sqrt{2\pi}}\int d n\frac{G( n)}{ n^{3/2}}\,e^{V S( n)} =
    \frac{1}{12\langle n\rangle V} G(\langle n\rangle)\,,
\end{equation}
\end{subequations}
So we have eventually
\begin{equation}\label{t lim 2}
    \langle G\rangle_V = G(\langle n\rangle) + \frac{\langle n\rangle}{2V}G^{''}(\langle n\rangle) +
    \mathcal{O}(V^{-2})\,,
\end{equation}
for any function $G$.

This gives us the exact expression for the density distribution
\eqref{probab dens} in the large volume limit
\begin{equation}\label{poison t lim 2}
    \mathcal{P}_{\langle n\rangle}(n;V)\sim\delta( n-\langle n\rangle)+\frac{\langle n\rangle}{2
    V}\,\delta^{''}( n-\langle n\rangle)+\mathcal{O}(V^{-2})\,.
\end{equation}

We are now able to obtain arbitrary density moments up to
$\mathcal{O}(V^{-2})$ terms.
\begin{equation}\label{moments}
    \langle n^k \rangle_V = \int dn\, n^k \mathcal{P}_{\langle
    n\rangle}(n;V) = \langle n\rangle^k +\frac{k(k-1)}{2V}\langle
    n\rangle^{k-1}+\mathcal{O}(V^{-2})\,.
\end{equation}

We have for the second moment (intensive variable!)

\[\langle n^2 \rangle_V = \langle n\rangle^2 + \frac{\langle n\rangle}{V}+\mathcal{O}(V^{-2})\,.\]

This means
\begin{equation}\label{density limit}
    \Delta n^2=\frac{\langle n\rangle}{V}\to 0\,.
\end{equation}
as expected in the thermodynamic limit.

The particle number and its density are fixed in the canonical ensemble so
corresponding variances are always equal to zero. The result \eqref{density limit}
can be seen as an example of the equivalence of the canonical and grand canonical
distribution in the thermodynamic limit. This equivalence is clearly visible from
the Eq \eqref{poison t lim 2} where the delta function in the first term can be
considered as the particle number density distribution in the canonical ensemble.

A more involved situation appears for particle number moments (extensive variable!).
Eq \eqref{moments} translated to the particle number gives
\begin{equation}\label{particle moments 2}
        \langle N^k\rangle = V^k\langle n\rangle^k + V^{k-1}\frac{k(k-1)}{2}\langle
        n\rangle^{k-1}+\mathcal{O}(V^{k-2})\,,
\end{equation}

 One gets for the scaled variance (semi-intensive variable!)
\begin{equation}\label{scaled variance}
    \frac{\Delta N^2}{\langle N\rangle}=1\,,
\end{equation}
what should be compared with zero obtained for the canonical distribution.

The mechanism for such a seemingly unexpected behavior is quite obvious. The grand
canonical and the canonical density probability distributions tend to the same
thermodynamic limit. There are different however for any finite volume.
Semi-intensive variables depend on coefficients at those finite volume terms so they
are different also in the thermodynamic limit.

\subsection{Energy distribution}
It is interesting to perform similar calculation for the energy
distribution in both ensembles. Energy moments and an average energy
density can be written as
\begin{equation}\label{energy moments}
    \langle E^k\rangle = (-1)^k\frac{1}{\mathcal{Z}}\frac{\partial^k\mathcal{Z}}
    {\partial\beta^k}\,;\qquad \langle \epsilon\rangle=-\frac{d z_0}{d\beta}\,e^\gamma\,.
\end{equation}

One gets from Eq \eqref{energy moments}
\begin{equation}\label{en moments GC-P}
    \langle E^k\rangle= V^k\langle\epsilon\rangle^k+V^{k-1}\frac{k(k-1)}{2}\langle\epsilon\rangle^{k-2}
    \frac{\langle n\rangle}{z_0}\frac{d^2 z_0}{d\beta^2}+\mathcal{O}(V^{k-2})\,.
\end{equation}
The grand canonical energy density distribution follows
\begin{equation}\label{energ probab}
    \mathbf{P}(\epsilon|\langle  n\rangle,\langle\epsilon\rangle)=
    \delta\left(\epsilon-\langle\epsilon\rangle\right)+
    \frac{\langle n\rangle}{2 V}\,
    \mathcal{R}^{GC}\left(\frac{\langle\epsilon\rangle}{\langle n\rangle}\right)
    \delta^{''}(\epsilon -\langle \epsilon\rangle)
    +\mathcal{O}(V^{-2})\,.
\end{equation}
$\mathcal{R}^{GC}$ is given here as
\[\mathcal{R}^{GC}\left(\frac{\langle\epsilon\rangle}{\langle n\rangle}\right)=
\left.\frac{1}{z_0}\frac{d^2
z_0}{d\beta^2}\right|_{\beta=\beta(\langle\epsilon\rangle/\langle n\rangle)}\,.\]

For the canonical distribution a corresponding statistical operator is
  \begin{equation}\label{stat operator C-P}
    \hat{D}=\frac{\,e^{-\beta\hat H}}{\,\text{Tr}{\,e^{-\beta\hat H}}}
\end{equation}

This leads to the partition function
\begin{equation}\label{C-P part fn}
    \mathcal{Z}(V,T)=\frac{z^N}{N!}=\frac{\,e^{Vn\log z}}{N!}\,.
\end{equation}

Internal energy moments are given by Eq \eqref{energy moments}. In
particular
\begin{equation}\label{av energy C-P}
    \langle \epsilon\rangle=-\frac{n}{z_0}\frac{d z_0}{d\beta}\,.
\end{equation}

For the energy moments one gets now
\begin{equation}\label{en moments C-P}
    \langle E^k\rangle=V^k\langle \epsilon\rangle^k +
    V^{k-1}\frac{k(k-1)}{2}\langle
    \epsilon\rangle^{k-2}n\frac{\partial}{\partial\beta}\left(\frac{1}{z_0}
    \frac{\partial z_0}{\partial\beta}\right)+\mathcal{O}(V^{k-2})\,.
\end{equation}

A corresponding probability distribution is
\begin{equation}\label{energ probab C-P}
    \mathbf{P}(\epsilon|n,\langle\epsilon\rangle)=
    \delta\left(\epsilon-\langle\epsilon\rangle\right)+
    \frac{n}{2 V}\,
    \mathcal{R}^{C}\left(\frac{\langle\epsilon\rangle}{n}\right)
    \delta^{''}(\epsilon -\langle \epsilon\rangle)
    +\mathcal{O}(V^{-2})\,,
\end{equation}
where $\mathcal{R}^{C}$ is given here as
\[\mathcal{R}^{C}\left(\frac{\langle\epsilon\rangle}{n}\right)=
\left.\frac{\partial}{\partial\beta}\left(\frac{1}{z_0}
    \frac{\partial z_0}{\partial\beta}\right)
    \right|_{\beta=\beta(\langle\epsilon\rangle/n)}\,.\]


\begin{thebibliography}{0}
\bibitem{brs} For a review see, \emph{e.g.}, P.~Braun-Munzinger, K.~Redlich and J.~Stachel: \emph{Quark Gluon Plasma 3}
eds. R.~C.~Hwa and X.~N.~Wang (World Scientific, Singapore 2004) 491-599;
A.~Andronic and P.~Braun-Munzinger: Lect.~Notes~Phys. \textbf{652} 35 (2004)
\bibitem{crt1} J.~Cleymans, K.~Redlich and L.~Turko: Phys.~Rev.~C \textbf{71} 047902 (2005)
\bibitem{crt2} J.~Cleymans, K.~Redlich and L.~Turko: J.~Phys.~G \textbf{31} 1421 (2005)
\bibitem{begun} V.~V.~Begun, M.~Gazdzicki, M.~I.~Gorenstein and O.~S.~Zozulya: Phys.~Rev.~C \textbf{
70} 034901 (2004); V.~V.~Begun, M.~I.~Gorenstein, A.~P.~Kostyuk and
O.~S.~Zozulya: Phys.~Rev.~C \textbf{ 71} 054904 (2005); V.~V.~Begun,
M.~I.~Gorenstein and O.~S.~Zozula: Phys.~Rev.~C \textbf{72} 014902
(2005); A.~Ker\"anen, F.~Becattini, V.V.~Begun, M.I.~Gorenstein,
O.S.~Zozulya, J.~Phys.~G \textbf{31} S1095 (2005)
\end{thebibliography}
\end{document}